\documentclass[hidelinks,11pt]{article}

\newcommand{\bol}{\boldsymbol}

\newcommand{\ney}{\boldsymbol{y}}                          
\newcommand{\nex}{\boldsymbol{x}}

\newcommand{\de}{\,\mathrm{d}}                               
\newcommand{\e}{\operatorname{e}}                               
\newcommand{\im}{\operatorname{i}}

\newcommand{\p}{\partial}

\newcommand{\real}{\mathrm{Re}\,}    
                                   
\newcommand{\imag}{\mathrm{Im}\,}

\usepackage{multirow}
\usepackage{amsmath}
\usepackage{amsfonts}
\usepackage{amsmath}
\usepackage{amssymb}  
\usepackage{graphicx}
\usepackage{caption}
\usepackage{mathrsfs}
\usepackage{upgreek}
\usepackage{amsthm}
\usepackage{subfig}
\usepackage{booktabs}
\usepackage{authblk}

\newtheorem{theorem}{Theorem}[section]

\newtheorem{remark}[theorem]{Remark}

\title{Modeling and simulation of an acoustic~well~stimulation~method}
\author[1]{Carlos P\'erez-Arancibia\footnote{E-mail: cperezar@mit.edu}}
\affil[1]{\small{Department of Mathematics, Massachusetts Institute of Technology}}
\author[2]{Eduardo Godoy}
\affil[2]{\small{INGMAT R\&D Centre, Jos\'e Miguel de la Barra 412, 4to piso, Santiago, Chile.}}
\author[2]{Mario Dur\'an}
\begin{document}
\maketitle

\begin{abstract}
This paper presents a mathematical model and a numerical procedure to simulate an acoustic well stimulation (AWS) method for enhancing the permeability of the rock formation surrounding oil and gas wells. The AWS method considered herein aims to exploit the well-known permeability-enhancing effect of mechanical vibrations in acoustically porous materials, by transmitting  time-harmonic sound waves from a sound source device---placed inside the well---to the well perforations made into the formation. The efficiency of the AWS is assessed by quantifying the amount of acoustic energy transmitted from the source device to the rock formation in terms of the emission frequency and the well configuration. A simple methodology to find \emph{optimal emission frequencies} for a given well configuration is presented. The proposed model is based on the Helmholtz equation and an impedance boundary condition that effectively accounts for the porous solid-fluid interaction at the interface between the rock formation and the well perforations. Exact non-reflecting boundary conditions derived from Dirichlet-to-Neumann maps are utilized to truncate the circular cylindrical waveguides considered in the model. The resulting boundary value problem is then numerically solved by means of the finite element method. A variety of numerical examples are presented in order to demonstrate the effectiveness of the proposed procedure for finding optimal emission frequencies.
\end{abstract}
\section{Introduction}\label{SC01:INTRODUCTION}
The decrease of oil and gas recovery from a reservoir is clearly an important problem that affects the energy industry. One of the main causes of such problem is the local reduction of the reservoir permeability around producing wells due to the deposition of scales, precipitants and mud penetration during exploitation which, over time, give rise to an impermeable barrier to fluid flow~\cite{GORBACHEV:1999}. Well stimulation methods play a prominent role in the exploitation of these essential natural resources as they are intended to increase the permeability of the reservoir, allowing the trapped fluid to flow toward the borehole and thus enhancing the productivity of the well. Various well stimulation methods are used in practice to cope with local deposits, including solvent and acid injection, treatment by mechanical scrapers and high pressure fracturing. Each one of these conventional methods have significant drawbacks and undesirable effects. Some of them, for instance, are expensive and produce damage to the well structure, while others are highly polluting, leading to harmful ecological effects associated with the contamination of underground water resources~\cite{GORBACHEV:1999,BERESNEV:1994}. The demonstrated effectiveness of mechanical vibrations on enhancing fluid flow through porous media~\cite{BERESNEV:1994,HAMIDA:2007,BATZLE:1992}, on the other hand, has led to the development of the so-called \emph{acoustic well stimulation}~(AWS) methods, which nowadays have broad acceptance by the hydrocarbon industry mainly due to the fact that they partially overcome the aforementioned issues.

This paper considers an AWS method based on the transmission of acoustic waves, emitted by a transducer submerged into the well, to the rock formation surrounding the well. The transducer is designed to trigger one of the physical processes known to enhance the permeability of the porous  medium. Among such physical processes, we mention the reduction of the fluid viscosity by agitation and heating, stimulation of elastic waves on the well walls (to reduce the adherence forces in the layer between oil and rock formation), excitation of natural frequencies associated with the vibration of the fluid inside the porous  medium,  and the formation and collapse of cavitation bubbles near clogged pores of the rock formation. A variety of transducer designs have been proposed over the last three decades, which consider operation frequency and intensity ranges selected to target one (or several) of the aforementioned physical processes~\cite{PECHKOV:1993,ELLINGSEN:1994,MAKI:1997,KOSTROV:2002}.


This paper presents a mathematical model and a numerical procedure that allows us to find optimal emission frequencies for which the amount of energy transmitted from the transducer into the rock formation is maximized. The proposed methodology can potentially improve the performance of the whole class AWS methods considered, as the aforementioned physical processes take place within the porous medium. In detail, we develop a mathematical model based  on the Helmholtz equation and an impedance boundary condition~\cite{FILIPPI:1999} that effectively accounts for the porous solid-fluid interaction at the interface between the rock formation and the well perforations~\cite{WHITE:1983}. Exact non-reflecting boundary conditions derived from Dirichlet-to-Neumann (DtN) maps are utilized to truncate the circular cylindrical waveguides considered in the model~\cite{GOLDSTEIN:1982,PEREZ:2010:1,PEREZ:2010}. The resulting boundary value problem is numerically solved by means of the finite element (FE) method~\cite{THOMSON:2006,IHLENBURG:1998,HARARI:2006:I}. Optimal emission frequencies are then found by scanning the quotient of the emitted energy to the transmitted energy---toward the region of interest---over a range of frequencies. As expected, the optimal emission frequencies correspond to field distributions for which resonances occur inside the perforations.

The outline of this paper is as follows: The mathematical model is presented in Section~\ref{SC02:MATH_MODEL}. The DtN-FE method is then described and validated in Section~\ref{SC03:DTNFEM}. Section~\ref{SC04:ACOUSTIC_WELL_RECOVERY} provides  numerical results for realistic well configurations. Section~\ref{SC05:CONCLUSIONS}, finally, gives the concluding remarks of the present work.

\begin{figure}[ht]
\begin{center}
 \includegraphics[scale=0.8]{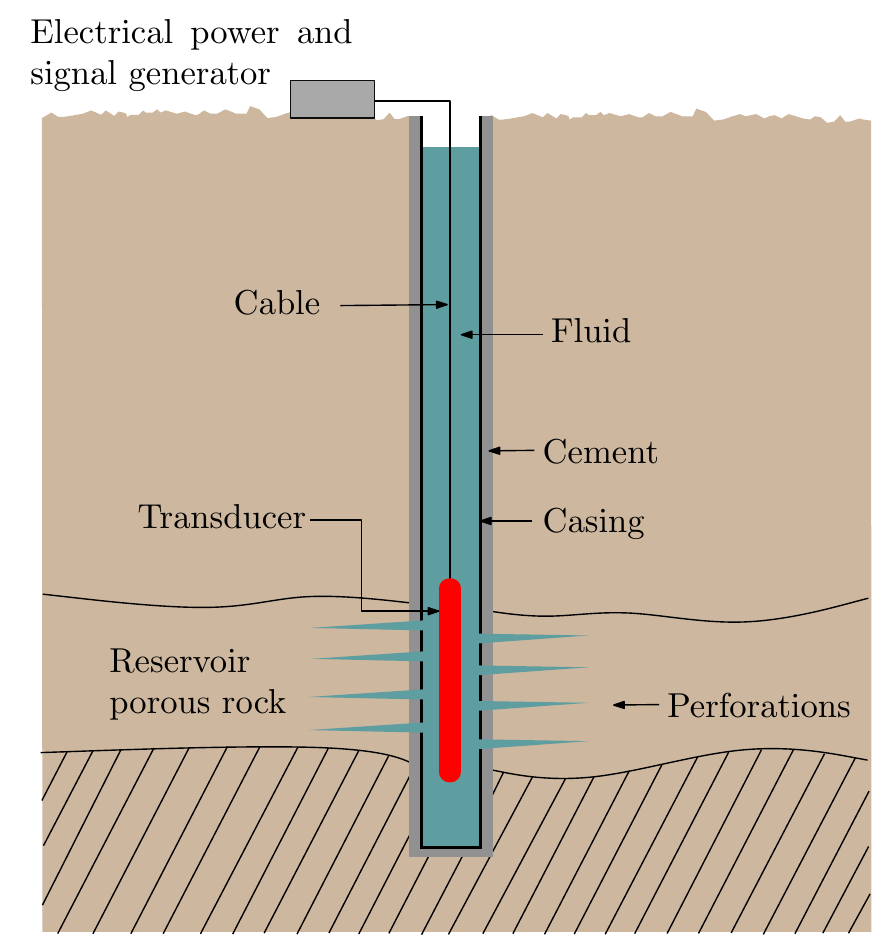}\\
 \caption{Diagram of the operation of an AWS method in a perforated (completed) well.}\label{FIG:SC01:AWS_SCHEME}
 \end{center}
\end{figure}


\section{Mathematical model}\label{SC02:MATH_MODEL}
\subsection{Geometry}\label{SSC:SC02:DOMAIN}
A perforated well is created through two successive processes called drilling and completion. The former begins by drilling a borehole in the ground, which is covered by metal pipes that are attached to its walls by a layer of cement (cf.~Figure \ref{FIG:SC01:AWS_SCHEME}). This part of the process,  commonly referred to as casing, aims to stabilize the borehole structure. Once the well is cased, the completion process begins by shooting with explosives the portion of the casing that passes through the reservoir level---where the oil is trapped---forming small holes across the casing and the cement layer, and into the reservoir. These holes, referred to as perforations, are aimed at enabling the oil to flow from the reservoir into the well.

\begin{figure}[ht]
\begin{center}
 \includegraphics[scale=0.55]{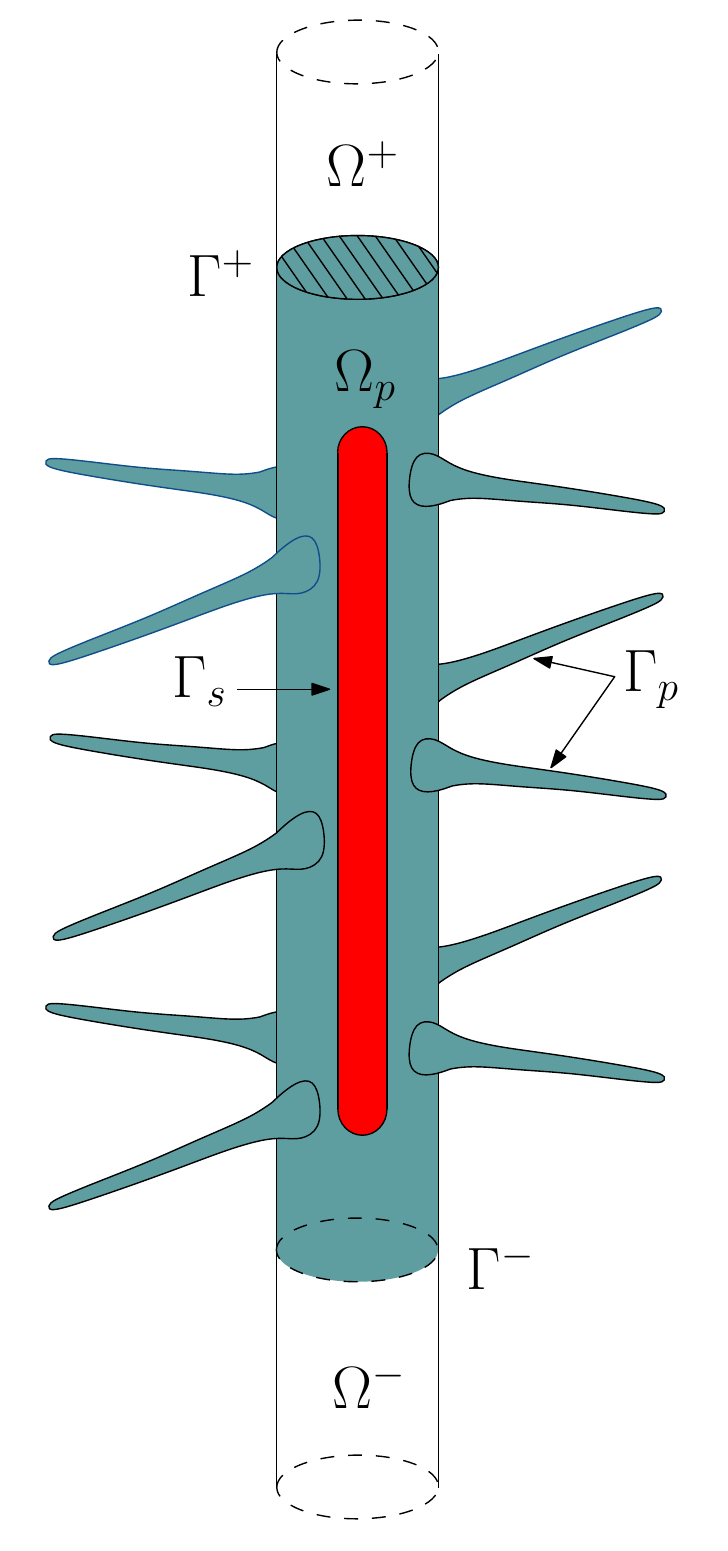}\\
 \caption{Geometric description of the perforated well and the transducer.}\label{FIG:SC02:WELL_GEOMETRY}
 \end{center}
\end{figure}

Upon completion, two different zones of the well can be identified; the zone containing the perforations, which we call the~\emph{perforated domain}, and the remaining part of the well, which we call the~\emph{cylindrical domain}. The perforated domain, denoted by~$\Omega_p$, is assumed to be bounded. In addition, we assume that the cylindrical domain consists  of two (semi-infinite) circular cylinders placed above and below the perforated domain, which we denote by $\Omega^+$ and~$\Omega^-$, respectively. The model of a perforated well utilized in this paper then, corresponds to a locally perturbed circular cylinder defined as $\Omega_w=\Omega_p\cup\Omega^+\cup\Omega^-$. The interface between the perforated and the upper (resp. lower) cylindrical domains is denoted by $\Gamma^+$ (resp. $\Gamma^-$). Finally, the transducer (source) is assumed to occupy the bounded domain $\Omega_s\subset\Omega_p$ with boundary $\p\Omega_s=\Gamma_s$. We refer to Figure~\ref{FIG:SC02:WELL_GEOMETRY} for the definition of all the relevant domains considered in the mathematical model. 

\subsection{Acoustic waves}
The transducer is herein modeled as a time-harmonic vibrating surface $\Gamma_s$ that operates at a fixed frequency $f=\omega/2\pi$, where $\omega>0$ denotes the angular frequency in radians.
Being excited by a single  time-harmonic source, the pressure $P$, the density $\varrho$, and the velocity  $\boldsymbol V$ fields  eventually reach a stationary (time-harmonic) regime for which $P(\nex,t)={\real}\left\{p(\nex)\e^{-i\omega t}\right\}$, $\varrho(\nex,t)={\real}\left\{\rho(\nex)\e^{-i\omega t}\right\}$, and  $\boldsymbol V(\nex,t)={\real}\left\{\boldsymbol v(\nex)\e^{-i\omega t}\right\}$, where $t>0$ denotes the time variable and $p$, $\rho$ and $\boldsymbol v$ denote the amplitudes of the pressure, the density and the velocity, respectively, which only depend on the position $\nex$.
The linearized equations of state and conservation of mass and momentum in this case, read as~\cite{FILIPPI:1999,KINSLER:1999}
\begin{subequations}
\begin{align}
 p &=c \rho,\label{eq:mat}\\
-\frac{i\omega}{c} p+\rho_0\,{\rm div}\:\boldsymbol v &= 0,\label{eq:mass}\\
-i\omega\boldsymbol v+\frac{1}{\rho_0}\nabla p &=0,\label{eq:mom}
\end{align}
\end{subequations}
where $c>0$ and $\rho_0>0$ denote the speed of sound and the equilibrium density of the fluid that fills the well, respectively. Suitably combining equations~\eqref{eq:mat},~\eqref{eq:mass} and~\eqref{eq:mom} we then obtain that $p$  satisfies the Helmholtz equation
\begin{equation}
\Delta p+k^2p=0\label{EQN:SC02:HELMHOLTZ}
\end{equation}
in the domain $\widetilde\Omega=\Omega_w\setminus\overline{\Omega_s}$ occupied by the fluid, where $k=\omega/c$ denotes the wavenumber.

Note that dissipation effects can be easily taken into account by considering a complex wavenumber with spatial absorption depending on the equilibrium density and the shear and bulk viscosities~\cite{KINSLER:1999}. For presentation simplicity, however, we only consider real wavenumbers.

\subsection{Boundary conditions}\label{SSC:SC02:BOUNDARY_CONDITIONS}
Throughout this paper we consider boundary conditions of the form
\begin{equation}
\frac{\p p}{\p n}-\frac{\im k}\zeta\,p=g\label{EQN:SC02:IMPEDANCE_BOUNDARY_CONDITION}
\end{equation}
on the surfaces of the well ($\Gamma_w=\p\Omega_w$) and on the transducer ($\Gamma_s$), where $\zeta\in\mathbb C$ denotes the dimensionless surface impedance and the function $g$ corresponds to the excitation prescribed on the surface $\Gamma_s$ of the transducer. The dimensionless impedance takes the form~$\zeta=\chi+\im \xi$, where $\chi$ and $\xi$ ($\chi,\xi:\Gamma_w\cup\Gamma_s\to\mathbb R$) are known as the resistive (real) and reactive (imaginary) parts of the impedance, respectively.  The dimensionless impedance $\zeta$ and the pressure field~$p$ are related to the time-averaged energy flux through $\Gamma_w$  by the formula~\cite{FILIPPI:1999}
\begin{equation}
I_{\mathrm{abs}}=\frac1{2\rho_0c}\int_{\Gamma_w}\frac{|p|^2}{|\zeta|^2}\chi\de s.\label{EQN:SC02:OUTFLOW_ENERGY}
\end{equation}
The time-averaged acoustic energy radiated by transducer, on the other hand, is given by
\begin{equation}
I_{\mathrm{rad}}=\frac1{2\rho_0\omega}\int_{\Gamma_s}{\imag}\left\{p\bar{g}\right\}\de s.
\label{EQN:SC02:INFLOW_ENERGY}
\end{equation}

The spatial dependence of the dimensionless impedance $\zeta$ in~\eqref{EQN:SC02:IMPEDANCE_BOUNDARY_CONDITION} is determined by the mechanical properties of the various materials that are in direct contact with the fluid. Being the casing made of metal (see Section~\ref{SSC:SC02:DOMAIN})---which is usually modeled as a sound hard (Neumann) boundary condition---the admittance $1/\zeta$ is taken equal to zero over the cylindrical domain and the cased portion of the perforated domain. The sound  hard boundary condition ($1/\zeta=0$) is also used on the transducer~$\Gamma_s$. In order to determine suitable impedance values to be used over the boundary of the perforations, in turn, we follow the analytical calculations presented by J.~E.~White in~\cite{WHITE:1983} for the wall impedance at the interface between a liquid and a porous material. According to these calculations, the wall impedance $Z$---defined as the quotient of the pressure amplitude to the normal velocity amplitude on the boundary of the perforation--- is given by
\begin{equation}
Z=\frac{p}{\boldsymbol v\cdot\boldsymbol n}=\Bigg(\frac{\kappa\sqrt{i\omega m}}{\eta}\frac{H^{(1)}_1(\sqrt{i\omega m}r_0)}{H^{(1)}_0(\sqrt{i\omega m}r_0)}\Bigg)^{-1},\label{eq:impedance:white}
\end{equation}
where $H^{(1)}_0$ and $H^{(1)}_1$ denote the Hankel functions of the first kind and order zero and one, respectively~\cite{ABRAMOVITZ:1972}, $r_0$ is the radius of the perforation, $\kappa$ is the permeability of the porous medium, $\eta$ is the shear viscosity of the fluid, and $m=\phi\eta/(\kappa B)$, being $\phi$ the porosity and $B$ the bulk modulus of the fluid in the pore space. It is important to highlight that the impedance model~\eqref{eq:impedance:white} is valid under the assumption that $r_0$ is smaller than the wavelength $\lambda=2\pi/k$. For the sake of completeness, the analytical derivations leading to~\eqref{eq:impedance:white} are reproduced in~\ref{app:impedance_bc}. On the other hand, in order to link $Z$ with the dimensionless surface impedance $\zeta$ we get,  from the momentum conservation equation~\eqref{eq:mom}, the relation
\begin{equation*}
\boldsymbol v\cdot\boldsymbol n=-\frac i{\omega\rho_0}\frac{\p p}{\p n},
\end{equation*}
which combined with the definition of $Z$ yields
\begin{equation}
\frac{\p p}{\p n}-\frac{ikc\rho_0}Z\,p=0.\label{eq:ibc:z}
\end{equation}
From \eqref{EQN:SC02:IMPEDANCE_BOUNDARY_CONDITION} with $g=0$ and \eqref{eq:ibc:z}, we obtain that $Z=\rho_0c\zeta$. Therefore, the dimensionless surface impedance to be utilized in \eqref{EQN:SC02:IMPEDANCE_BOUNDARY_CONDITION} on the surface of the perforations is given by
\begin{equation}
\zeta=\Bigg(\frac{\rho_0c\kappa\sqrt{i\omega m}}{\eta}\frac{H^{(1)}_1(\sqrt{i\omega m}r_0)}{H^{(1)}_0(\sqrt{i\omega m}r_0)}\Bigg)^{-1}.\label{eq:impedance}
\end{equation}

\subsection{Boundary value problem}\label{SSC:SC02:SET-UP}
We are now in position to put together the boundary value problem to be solved in what follows of this paper.  The time-harmonic pressure field $p:\widetilde\Omega\to\mathbb{C}$, which is driven by the transducer submerged into the well, satisfies
\begin{subequations}\begin{eqnarray}
\Delta p+k^2p&=&0\quad\text{in}\quad\widetilde \Omega,\medskip\label{eq:HLM}\\
\frac{\p p}{\p n}-\frac{ik}\zeta\,p&=&0\quad\text{on}\quad\Gamma_w,\label{eq:NBC}\medskip\\
\frac{\p p}{\p n}&=&g\quad\text{on}\quad\Gamma_s,
\end{eqnarray}\label{EQN:SC02:SCATTERING_PROBLEM}\end{subequations}
where the dimensionless impedance $\zeta$ is given by~\eqref{eq:impedance} on the boundary of perforations and it equals infinity (i.e., $1/\zeta=0$) everywhere else on $\Gamma_w$ (see Section~\ref{SSC:SC02:BOUNDARY_CONDITIONS}). In order for the boundary value problem~\eqref{EQN:SC02:SCATTERING_PROBLEM} to be well-posed,  $p$ has to  satisfy a certain radiation condition---which differs from the classical Sommerfeld condition---that is expressed in terms of the propagative modes associated with the upper and lower unbounded cylindrical domains $\Omega^+$ and $\Omega^-$~\cite{GOLDSTEIN:1982,PEREZ:2010:1,PEREZ:2010}.

\section{Dirichlet-to-Neumann Finite Element Method}\label{SC03:DTNFEM}

\subsection{The DtN map}
In what follows we present a DtN-FE method for the numerical solution of~\eqref{EQN:SC02:SCATTERING_PROBLEM}. Notice that standard finite element (FE) methods do not directly apply to this problem due to the unboundedness of the domain~$\widetilde\Omega$.
The DtN-FE method is based on the DtN operators $\mathcal T^\pm$ that map the boundary values $p|_{\Gamma^\pm}$ on $\Gamma^\pm$ into the corresponding normal derivatives~$\p p/\p n|_{\Gamma^\pm}$ on $\Gamma^\pm$~\cite{GOLDSTEIN:1982,PEREZ:2010:1,BENDALI:1999}.  As these DtN maps provide exact non-reflecting boundary conditions on $\Gamma^\pm$ they allow us to write a boundary value problem posed on the bounded domain $\Omega=\widetilde\Omega\setminus\overline{(\Omega^+\cup\Omega^-)}=\Omega_p\setminus\overline\Omega_s$ that is equivalent to~\eqref{EQN:SC02:SCATTERING_PROBLEM} and is suitable to be solved by FE methods (or any other standard numerical method for solving PDEs).

In order to provide explicit expressions for the DtN maps, we first introduce a cylindrical coordinate system $(r,\theta,z)$, with $r\geq 0, 0\leq\theta\leq 2\pi$ and $z\in\mathbb{R}$, upon which the upper and lower cylindrical domains can be expressed as $\Omega^\pm = \{r<R,\pm z>H \}\subset\mathbb{R}^3$, where $H>0$ denotes the truncation height and $R>0$ denotes the radius of the well. The series representation of the desired DtN maps are then obtained by applying the method of separation of variables to solve the Helmholtz equation in the domains $\Omega^\pm$ with Neumann boundary condition on the surface $\{r=R\}$. Enforcing the radiation condition---by eliminating both down-going (resp. up-going) and exponentially growing solutions in $\Omega^+$ (resp. $\Omega^-$)---we obtain the following Fourier-Bessel series for the pressure field~\cite{PEREZ:2010:1} 
\begin{equation}
p(r,\theta,z) = \sum_{n=-\infty}^\infty\sum_{m= 1}^\infty p^\pm_{n,m}v_{n,m}(r,\theta)\e^{\pm i(z\mp H)\sqrt{k^2-\lambda_{n,m}^2}}\quad\mbox{in}\quad\Omega^\pm, \label{eq:bessel_series}
\end{equation}
where, letting $j'_{n,m}\geq 0$ denote the $m$-th non-negative zero of the derivative of the Bessel function of first kind $J_n$, we have that
$$v_{n,m}(r,\theta) = c_{n,m}J_{n}\left(\lambda_{n,m}r \right)\e^{in\theta}\quad\mbox{and}\quad \lambda_{n,m}=\frac{j'_{n,m}}{R},
$$
with
$$
c_{n,m} = \left\{\begin{array}{ccc}\displaystyle\frac{\lambda_{n,m}}{\sqrt{2\pi}\sqrt{\lambda^2_{n,m}R^2-n^2}J_n(\lambda_{n,m}R)}&\mbox{if}&\lambda_{n,m}>0,\medskip\\
\displaystyle\frac{1}{\sqrt{2\pi}R}&\mbox{if}&\lambda_{n,m}=0,\end{array}\right.
$$
correspond to  the normalized Neumann-Laplace eigenfunctions and eigenvalues of the circle $\{r<R\}\subset\mathbb{R}^2$, respectively (i.e., they satisfy
$$\Delta v_{n,m}+\lambda_{n,m}^2 v_{n,m}=0\quad {\rm in }\quad \{r<R\},\quad \frac{\p v_{n,m}}{\p n} =0\quad {\rm on}\quad \{r=R\},\quad\mbox{and}\quad\int_{\{r<R\}} |v_{n,m}|^2=1.)$$
 The Fourier coefficients $p^\pm_{n,m}$ in~\eqref{eq:bessel_series}, in turn,  are given by
 \begin{equation*}
p^\pm_{n,m}=\int_0^R\!\!\int_{0}^{2\pi}p(r,\theta,\pm H)\overline{v_{n,m}(r,\theta)}r\de\theta\de r,\quad -\infty<n<\infty,\quad m\geq 1,
\end{equation*}
where  $p(r,\theta,\pm H)=p|_{\Gamma^\pm}$. Taking normal derivative of~\eqref{eq:bessel_series} on $\Gamma^\pm$ (with unit normal vectors pointing toward $\Omega^\pm$) we finally arrive at the following expression for the DtN maps
 \begin{equation}
\begin{split}
\mathcal T^{\pm} \left[\,p\,\right](\nex)=\displaystyle\sum_{n=-\infty}^\infty\sum_{m= 1}^\infty i\sqrt{k^2-\lambda_{n,m}^2}\, p^\pm_{n,m} v_{n,m}(r,\theta),\qquad \nex=(r\cos\theta,r\sin\theta,\pm H)\in\Gamma^{\pm}.
\end{split}\label{EQN:SC03:DTN_MAP_SUP}
\end{equation}

\subsection{Equivalent boundary value problem}
Using the continuity of the pressure field and its normal derivative across $\Gamma^\pm$ we thus obtain the following equivalent boundary value problem
\begin{subequations}\begin{eqnarray}
\Delta p+k^2 p&=&0\quad\text{in}\quad\Omega,\label{EQN:SC03:Helmholtz}\medskip\\
\displaystyle\frac{\p p}{\p n}-\frac{ik}\zeta\,p&=&0\quad\text{on}\quad\Gamma_p,\medskip\\
\displaystyle\frac{\p p}{\p n}&=&g\quad\text{on}\quad\Gamma_s,\medskip\\
\displaystyle\frac{\p p}{\p n}&=&\mathcal T^{\pm}\,p\quad\text{on}\quad\Gamma^{\pm},
\end{eqnarray}\label{EQN:SC03:TRUNCATED_SCATTERING_PROBLEM}\end{subequations}
for the pressure field in the bounded domain $\Omega$.

Multiplying the Helmholtz equation~\eqref{EQN:SC03:Helmholtz} across by a test function $q\in H^1(\Omega) $ and integrating by parts, we arrive at the variational (or weak) formulation of~\eqref{EQN:SC03:TRUNCATED_SCATTERING_PROBLEM}, which is expressed as follows: Find  $p\in H^{1}(\Omega)$ such that
\begin{equation}
a(p,q)=f(q), \qquad \forall q\in
H^1(\Omega),
\label{EQN:SC03:VARIATIONAL_FORMULATION_SCATTERING}
\end{equation}
where
\begin{subequations}\label{EQN:SC03:BILINEAR_FORM_AND_FUNCTIONAL}
\begin{align}
a(p,q)&=\int_{\Omega}\left(k^2\overline{q}\,p-\nabla\overline{q}\cdot\nabla p\right)\de \nex
+\int_{\Gamma_p} \frac{ik}\zeta\overline{q}\,p\de s
+\int_{\Gamma^+}\overline{q}\,\mathcal T^+p\de s
+\int_{\Gamma^-}\overline{q}\,\mathcal T^-p\de s, \label{EQN:SC03:BILINEAR_FORM}\\
f(q)&=-\int_{\Gamma_s} \overline{q}g\de s.
\end{align}
\end{subequations}
The well-posedness of the variational problem~\eqref{EQN:SC03:VARIATIONAL_FORMULATION_SCATTERING} can be easily established following the analysis presented in~\cite{GOLDSTEIN:1982}.
\subsection{Finite element discretization}

The discretization of the variational formulation~\eqref{EQN:SC03:BILINEAR_FORM_AND_FUNCTIONAL} by finite elements is straightforward. We consider a family of regular tetrahedral meshes  $\mathcal{T}_h$ of the domain $\Omega$, such that $\overline{\Omega}=\bigcup_{T\in\mathcal{T}_h}T$  ($\Omega$ is assumed to be a tetrahedral domain) where $h=\max\{\mathrm{diam}\,T: T\in\mathcal{T}_h\}$, with $\mathrm{diam}\,T=\max\{|\nex_1-\nex_2|:\nex_1,\nex_2\in T\}$.
Using  standard linear Lagrange elements, the approximate solution $p_h$ of~\eqref{EQN:SC03:BILINEAR_FORM_AND_FUNCTIONAL} is  expressed as
\begin{equation}
p_h(\nex)=\sum_{i=1}^{N}p_i\,\phi_i(\nex),\quad \nex\in\Omega,\label{EQN:SC03:FEM_EXPANSSION_SOLUTION}
\end{equation}
where $N$ is the number of nodes of the mesh and $\{\phi_1,\phi_2,\ldots,\phi_N\}$ is the nodal basis of the finite dimensional function space
$V_h=\big\{q\in H^1(\Omega):\ q\in\mathcal{C}^0(\Omega),\ q\mid_{T}\in \mathcal{P}_1(T),\ \forall\,T\in\mathcal{T}_h\big\}\subset H^1(\Omega)$ where $\mathcal{C}^0(\Omega)$ denotes the set of continuous functions in $\Omega$, and $\mathcal{P}_1(T)$ denotes the set of polynomials of degree at most one defined in $T$.
A system of equations for the node values $p_i$, $i=1,\ldots,N$ in~\eqref{EQN:SC03:FEM_EXPANSSION_SOLUTION} is obtained by substituting $p$ by $p_h$ in~\eqref{EQN:SC03:BILINEAR_FORM_AND_FUNCTIONAL} and taking test functions $q_h$ from the nodal basis of $V_h$. Doing so, and further replacing the bilinear form $a$ by an approximate bilinear form $\tilde a$, given by~\eqref{EQN:SC03:BILINEAR_FORM} but with the DtN maps $\mathcal T^\pm$ in the last two integrals expressed in terms of truncated series representations, we obtain the linear system
$$A\bol{p}=\bol{f},$$
where $A_{ij}=\tilde a(\phi_i,\phi_j)$, $1\leq i,j\leq N$, $\bol p=[p_1,\ldots,p_N]^T$ and $\bol f=[f(\phi_1),\ldots,f(\phi_N)]^T$.
In order to ensure the uniqueness of the solution of the linear system, it suffices to consider truncated series representations of the DtN maps that include all  the modes  satisfying $|\lambda_{n,m}|\leq k$~\cite{HARARI:1998}.

\begin{remark}
It is worth mentioning that one of the main advantages of the proposed absorbing boundary conditions over perfectly matched layers (PMLs) lies in the fact that the absorbing boundaries $\Gamma^\pm$ can be placed arbitrarily close to the region of interest (near the perforations and the transducer) provided that a sufficiently large number of modes are considered in the truncated series representations of the DtN maps. Off-the-shelf PMLs that absorb only propagative modes, on the other hand, would have to be placed far away enough from the region of interest so that all the evanescent modes are sufficiently attenuated, leading to larger computational domains and larger linear systems. Alternatively, PMLs that absorb both propagative and evanescent modes can also be used, provided that the mesh is properly refined to account for the frequency increment within the absorbing layers~\cite{johnson2008notes}.
\end{remark}

\subsection{Validation}\label{sec:val}
In this section, we present a numerical experiment  devised to validate the proposed DtN-FE method. We thus consider a test geometry consisting of a non-perforated well and a spherical transducer, given by  $\Omega_w=\left\{\nex=(r\cos\theta,r\sin\theta,z)\in\mathbb{R}^3: r<R\right\}\subset\mathbb{R}^3$ and $\Omega_s=\left\{\nex\in\mathbb{R}^3:|\nex-\ney|<\delta\right\}$, respectively, where $\Omega_s$ is centered at a point $\ney\in\Omega_w$ and $\delta>0$ is small enough so that  $\overline{\Omega_s}\subset\Omega_w$. On the spherical surface of the transducer, we prescribe the excitation
\begin{equation}
g(\nex)=\frac{\p G}{\p n_{\nex}}(\nex,\ney),\quad\nex\in\Gamma_s,\label{EQN:SC03:BENCHMARK_BOUNDARY_DATA}
\end{equation}
where $G$ is the Green's function of the infinite cylinder with homogeneous Neumann boundary conditions, which can be expressed in terms of the Neumann-Laplace eigenfunctions~\cite{POLYANIN:2002} as
\begin{equation*}
G(\nex,\ney)=\sum_{n=-\infty}^{\infty}\sum_{m=1}^\infty\frac{v_{n,m}(r,\theta)\overline{v_{n,m}(\rho,\vartheta)}}{\sqrt{\lambda_{n,m}^2-k^2}}\e^{-\sqrt{\lambda^2_{n,m}-k^2}|z-\zeta|},
\end{equation*}
with $\nex=(r\cos\theta,r\sin\theta,z)$ and $\ney=(\rho\cos\vartheta,\rho\sin\vartheta,\zeta)$.
It is easy to verify, from the definition of the Green's function, that
 \begin{equation}
 p(\nex)=G(\nex,\ney),\quad \nex\in\widetilde\Omega=\Omega_w\setminus\overline\Omega_s, \label{eq:exact_sol}
 \end{equation}
is in fact the exact solution of~\eqref{EQN:SC02:SCATTERING_PROBLEM} for the test geometry considered. This exact solution~\eqref{eq:exact_sol} is then compared with  approximate solutions obtained by means of the DtN-FE method described in Section~\ref{SC03:DTNFEM} for various mesh sizes $h>0$. In order to compare both the exact and the approximate solution, we define the relative error
\begin{equation}
\displaystyle E_h=\frac{\|p_h-\Pi_h p\|_{L^2\left(\Omega\right)}}{\|\Pi_h
p\|_{L^2\left(\Omega\right)}},\label{eq:rel_errors}
\end{equation}
where $\Pi_h p$ denotes the Lagrange interpolation of the exact solution using the tetrahedral  mesh $\mathcal T_h$.

\begin{figure}[h]
\begin{center}
\includegraphics[scale=0.5]{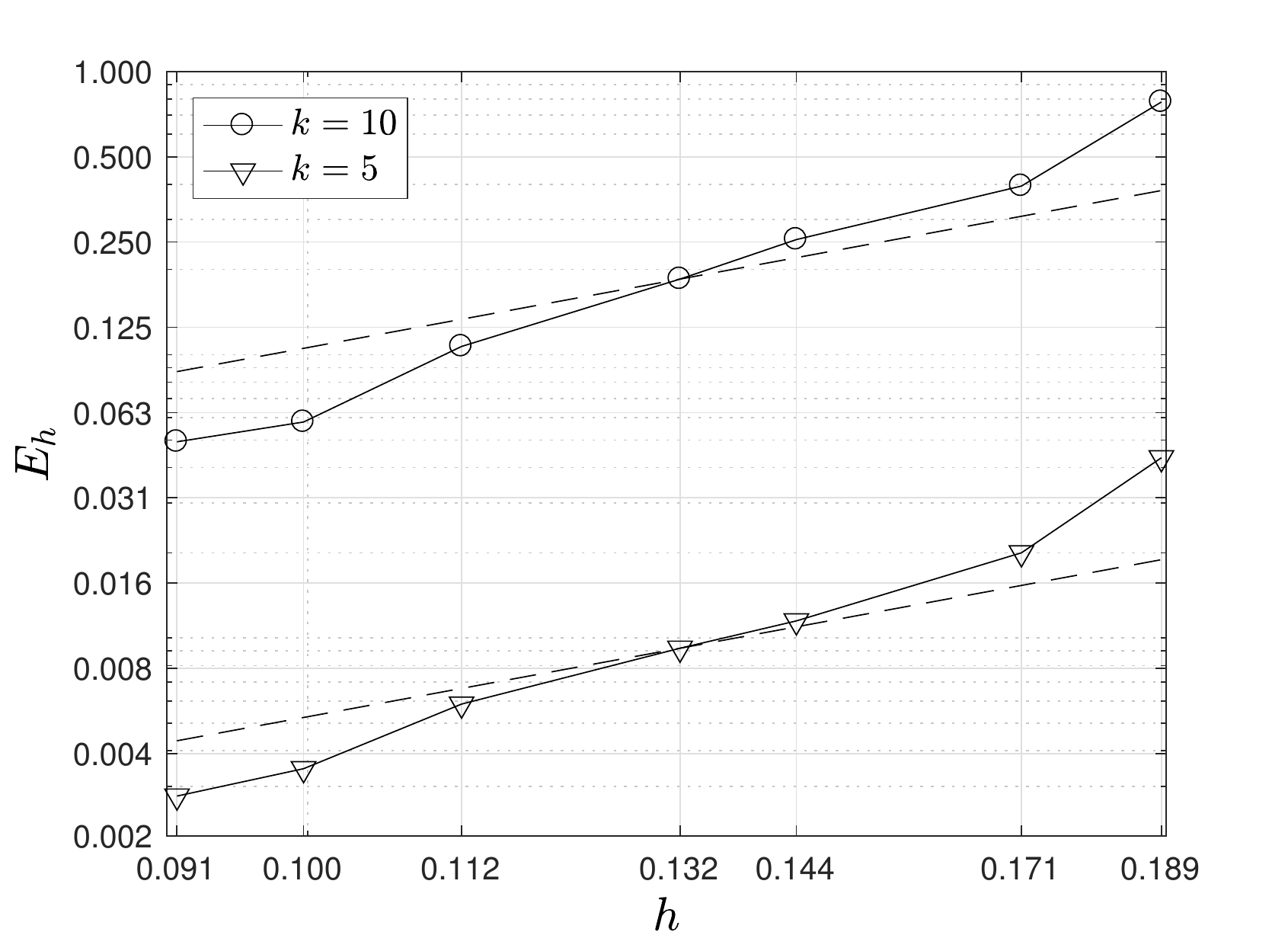}\\
\caption{Relative errors~\eqref{eq:rel_errors} in log-log scale in the solution of the test problem presented in Section~\ref{sec:val}, for various mesh sizes $h>0$ and wavenumbers. The dashed lines indicate second-order slopes.}\label{FIG:SC03:RELATIVE_ERROR_SCATTERING_NON_AXISYMMETRIC}
\end{center}
\end{figure}

The results of this numerical experiment are presented in Figure \ref{FIG:SC03:RELATIVE_ERROR_SCATTERING_NON_AXISYMMETRIC}, which displays the relative numerical errors~\eqref{eq:rel_errors} for the test problem with $R=0.5$, $\ney=(0,0.25,0)$ and $\delta=0.2$. The unbounded computational domain $\widetilde \Omega$  was truncated by introducing artificial boundaries~$\Gamma^\pm$ placed at $z=\pm H$, with $H=1.5$. Clearly, the numerical solution converges to the exact solution as the grid size tends to zero at a rate that is slightly faster than the expected second-order rate.

\section{Numerical simulations}
\label{SC04:ACOUSTIC_WELL_RECOVERY}
This section presents numerical simulations of the AWS method modeled in this paper. The values of the relevant physical constants of the fluid and the porous material---needed to evaluate the wavenumber $k=\omega/c$ and the surface impedance $\zeta$ in~\eqref{eq:impedance}---are displayed in Table~\ref{TABLE:SC04:EXAMPLE_PHYSICAL_CONSTANTS}. In detail, the fluid is assumed to be crude oil, with physical constants taken from~\cite{BATZLE:1992}, and the porous rock formation is assumed to be sandstone, with permeability and porosity values obtained from~\cite{WHITE:1988}.

In order to properly simulate the operation of the AWS method, the excitation $g$ on the surface of the transducer has to be suitably prescribed. For that purpose, the transducer is modeled as a constant-amplitude time-harmonic vibrating surface $\Gamma_s$ with $g=1\ \mathrm{N\,m}$. A more sophisticated  transducer model can be easily incorporated into the simulations by considering more general functions $g\in H^{-1/2}(\Gamma_s)$.

\begin{figure}[ht]
\begin{center}
\includegraphics[width=10cm]{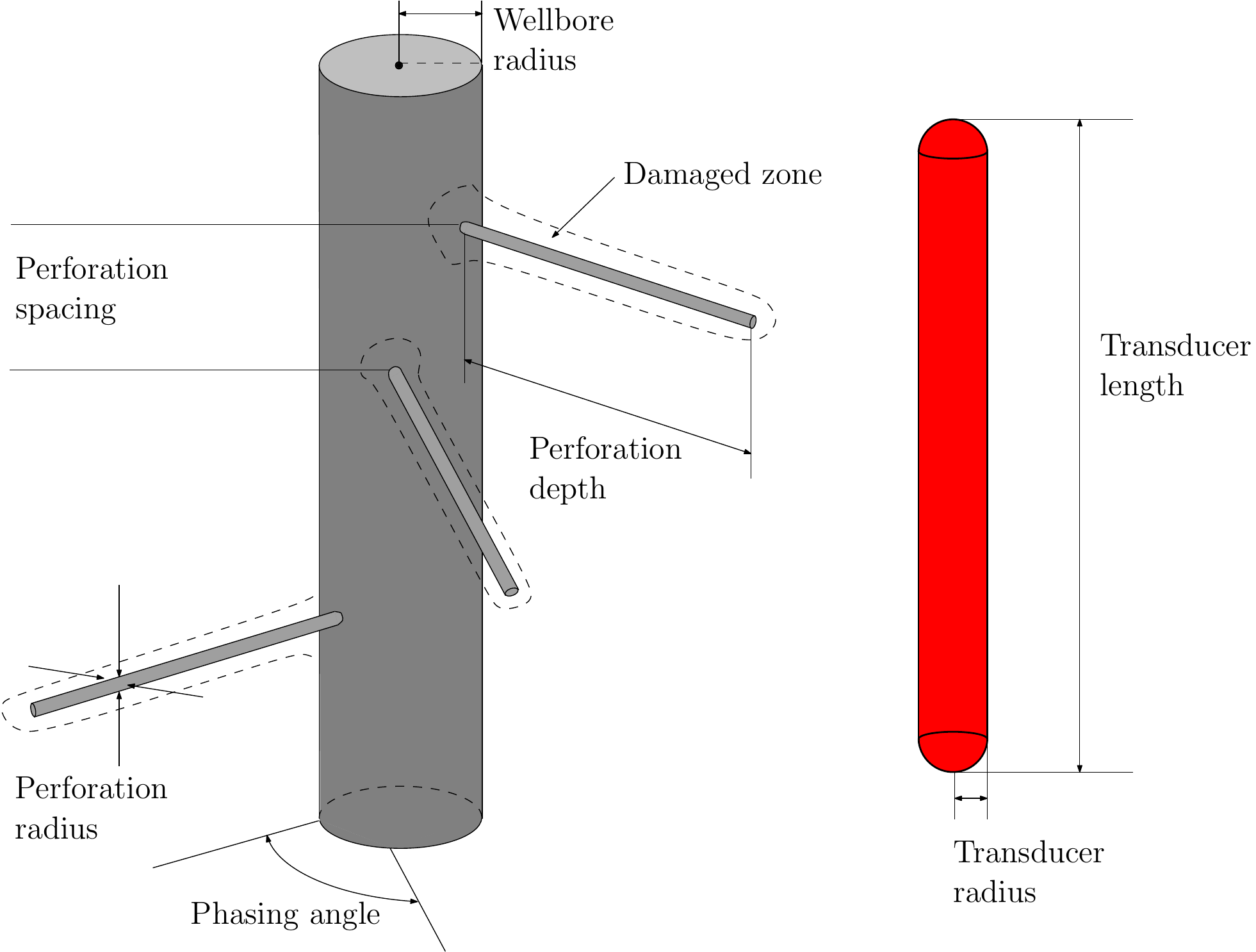}\\
\caption{Geometrical parameters utilized in the definition of the realistic well configuration and transducer.}\label{FIG:SC04:WELL_GEOMETRY_PARAMETERS}
\end{center}
\end{figure}

\begin{figure}[h!]
\centering	
 \subfloat[][1st well.]{\includegraphics[height=8cm]{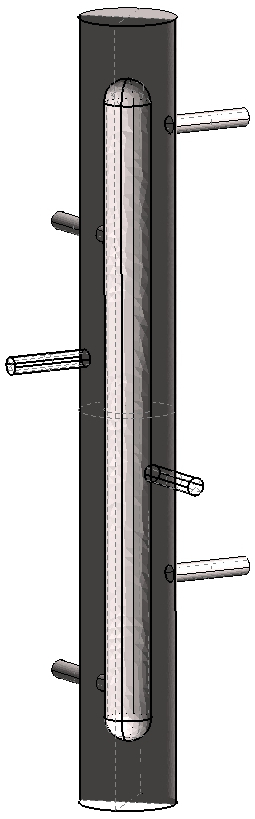}}\qquad
 \subfloat[][2nd well.]{\includegraphics[height=8cm]{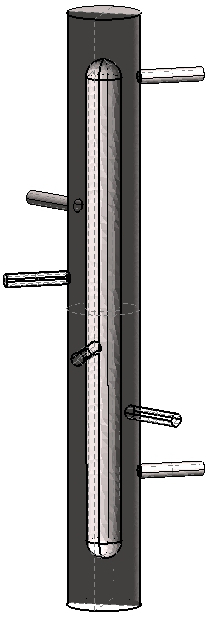}}\qquad
  \subfloat[][3rd well.]{\includegraphics[height=8cm]{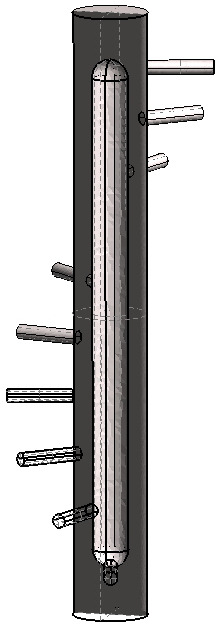}}\qquad
\caption{Well configurations considered in the numerical simulations.}\label{FIG:CH06:WELL_MESH}
\end{figure}

The generic well configuration to be considered in the simulations is depicted in~Figure~\ref{FIG:SC04:WELL_GEOMETRY_PARAMETERS}, which includes the definition of  the  relevant geometrical parameters. Three particular well configurations are initially considered, with specific geometrical  parameters provided in Table~\ref{TABLE:SC04:EXAMPLE_GEOMETRICAL_DATA}. The 1st, 2nd and 3rd well configurations include $N_p=6,8$ and $10$ perforations, respectively. The resulting computational domains, which were meshed using {Gmsh}~\cite{GEUZAINE:2009}, are shown in~Figure~\ref{FIG:CH06:WELL_MESH}.

\begin{table}[h]
\caption{Physical constants for crude oil and sandstone. The numerical values of the fluid constants were taken from~\cite{CHENG:2008}. The numerical values of the porous solid constants, on the other hand, were taken from~\cite{WHITE:1983}.}\vspace{0.2cm} \centering
\begin{tabular}{ll}
\hline \hline
Constant&Value \\
\hline
Speed of sound in oil ($c$) & 1524 m$\,$s$^{-1}$\\
Oil density ($\rho_0$) &1100 kg$\,$m$^{-3}$\\
Oil shear viscosity ($\eta$) & 1.2 Pa$\,$s\\
Oil bulk modulus ($B$) & 3000 MPa\\
Rock formation permeability ($\kappa$) & $3\times10^{-13}$ m$^2$\\
Rock formation porosity ($\phi$) & $0.21$\\
\hline \hline
\end{tabular}
\label{TABLE:SC04:EXAMPLE_PHYSICAL_CONSTANTS}
\end{table}

\begin{table}[h!]
\caption{Geometrical parameters of the well configurations considered. The dimensions of the transducer were selected according the device described in~\cite{MULLAKAEV:2009}. The  dimensions of a perforated well, on the other hand,  were taken from~\cite{HAGOORT:2007}}\vspace{0.2cm} \centering
\begin{tabular}{ll}
\hline \hline
Parameter&Value \\
\hline
Well radius ($R$) & 0.111 m \\
Perforated domain height ($2H$) &1.800 m\\
Transducer length & 1.410 m\\
Transducer radius & 0.054 m\\
Perforation radius ($r_0$)& 0.020 m \\
Perforation depth & 0.305 m \\\hline
{Perforation spacing (1st well)} & {0.257 m} \\
{Phasing angle (1st well)} &{$\pi/2$ rad}  \\\hline
{Perforation spacing (2nd well)} & {0.200 m} \\
{Phasing angle (2nd well)} &{$\pi/3$ rad}  \\\hline
{Perforation spacing (3rd well)} & {0.160 m} \\
{Phasing angle (3rd well)} &{$\pi/6$ rad}  \\
\hline \hline
\end{tabular}
\label{TABLE:SC04:EXAMPLE_GEOMETRICAL_DATA}
\end{table}

%
%

Next, we compute the energy transmission through the surface of the perforations and the energy emitted by the transducer using formulae \eqref{EQN:SC02:OUTFLOW_ENERGY} and \eqref{EQN:SC02:INFLOW_ENERGY}, respectively, for a certain range of frequencies $f=\omega/(2\pi)$. In order to find (local) optimal emission frequencies, we look for local maxima of the individual and consolidated energy transmission factors, that is,
\begin{subequations}
\begin{align}
Q_j&=\frac{I^j_\mathrm{abs}}{I_\mathrm{rad}}=\displaystyle k\chi \frac{\displaystyle\int_{\Gamma_p^j}\frac{\left|p\right|^2}{|\zeta|^2}\de s}
{\displaystyle\int_{\Gamma_s }\imag\left\{p\overline{g}\right\}\de s},
\hspace{0.5cm}j=1,\ldots,N_p,\quad \mbox{and}\label{eq:Q_j_factor}\\
Q&=\frac{I_\mathrm{abs}}{I_\mathrm{rad}}=\sum_{j=1}^{N_p}Q_j, \label{eq:Q_factor}
\end{align}\label{EQN:SC04:PROPORTION_PERFORATION}
\end{subequations}
respectively---which are dimensionless quantities---as functions of the excitation frequency $f=\omega/(2\pi)$. The pressure field $p$ in \eqref{eq:Q_j_factor} and \eqref{eq:Q_factor} corresponds to the solution of~\eqref{EQN:SC03:TRUNCATED_SCATTERING_PROBLEM} and $\Gamma_p^j\subset\Gamma_p$, $j=1,\cdots,N_p$, denotes the surface of the $j$-th perforation (the perforations are sorted from top to bottom).  Note that in virtue of the conservation of energy principle and the fact that $0\leq Q\leq 1$, we have that the quantity $100\times Q$ corresponds to the percent of energy effectively transmitted to the porous reservoir rock through the perforations.

\begin{figure}[h!]
\begin{center}
\includegraphics[height=4.5cm]{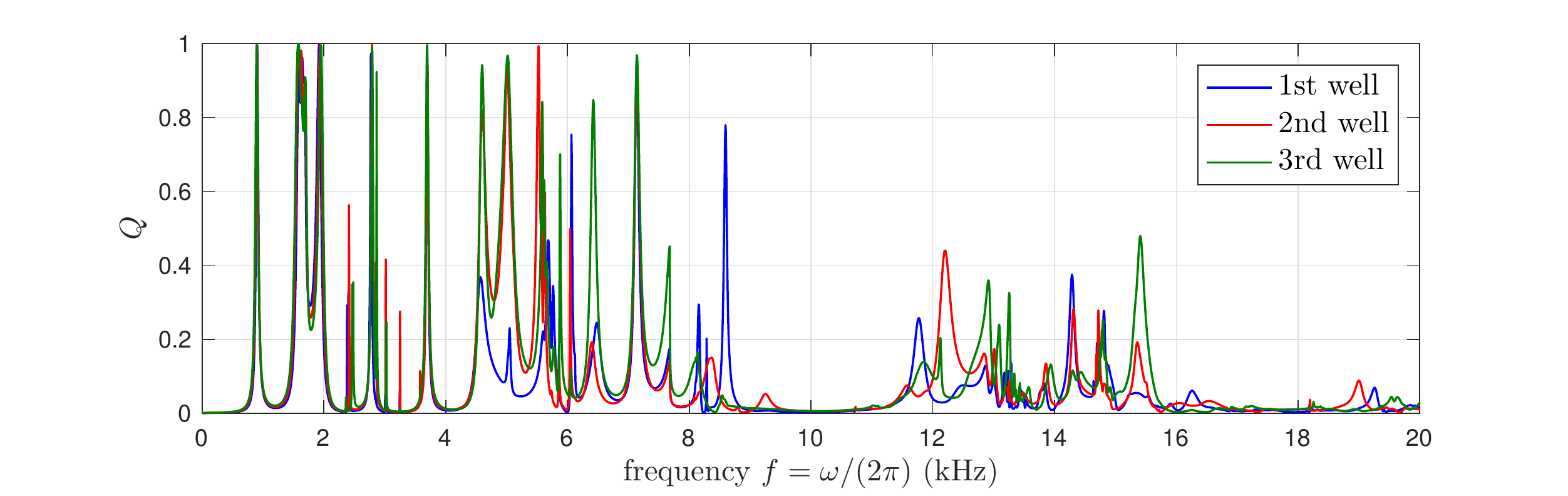}
\end{center}\vspace{-0.5cm}
\caption{Consolidated energy transmission factor $Q$~\eqref{eq:Q_factor} as a function of the frequency for the three ``symmetric" well configurations displayed in  Figure~\ref{FIG:CH06:WELL_MESH}.}\label{fig:Q_factor}
\end{figure}

\begin{figure}[h!]
\begin{center}
\includegraphics[width=10.0cm]{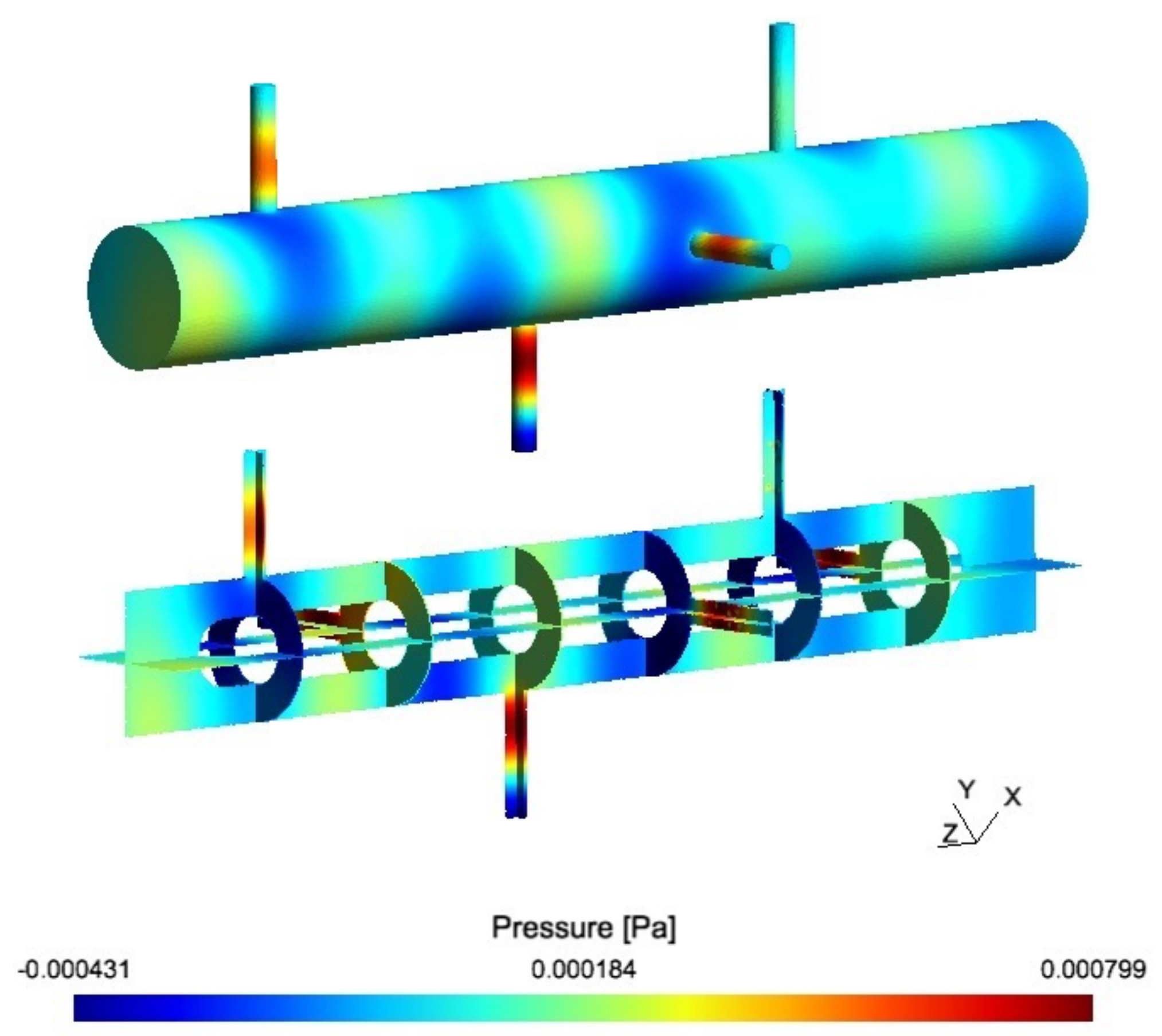}
\end{center}\vspace{-0.5cm}
\caption{Real part of the pressure field inside the 1st well configuration at $f=3.6941$ kHz, which corresponds to one of the peak values of the $Q$ factor displayed in Figure~\ref{fig:Q_factor} in blue. Top: pressure field on the boundaries of the computational domain $\Omega$. Bottom: pressure field at various cross sections of $\Omega$.}\label{fig:res_1}
\end{figure}

Figure~\ref{fig:Q_factor} displays the consolidated energy transmission factor $Q$ as a function of the excitation frequency~$f=\omega/(2\pi)$ for the three well configurations laid out in Table~\ref{TABLE:SC04:EXAMPLE_GEOMETRICAL_DATA} and Figure~\ref{FIG:CH06:WELL_MESH}. In these numerical simulations, the transducer is placed exactly at the center of the perforated domain.  Sharp peaks of the energy transmission factor $Q$---many of them reaching values close to the upper bound $Q=1$---are observed at various frequencies for the  well configurations considered (e.g., the peaks values around $f=0.895, 1.585, 2.79, 3.695$ and 5.525 kHz). The existence of these peaks is explained by resonance phenomena taking place inside the perforations. As illustrated by the pressure field at a peak frequency displayed in Figure~\ref{fig:res_1}, the factor $Q$ attains its local maxima at ``resonance" frequencies, for which the associated pressure field exhibits inordinate large amplitudes inside the perforations.

\begin{figure}[h!]
\begin{center}
\includegraphics[height=4.5cm]{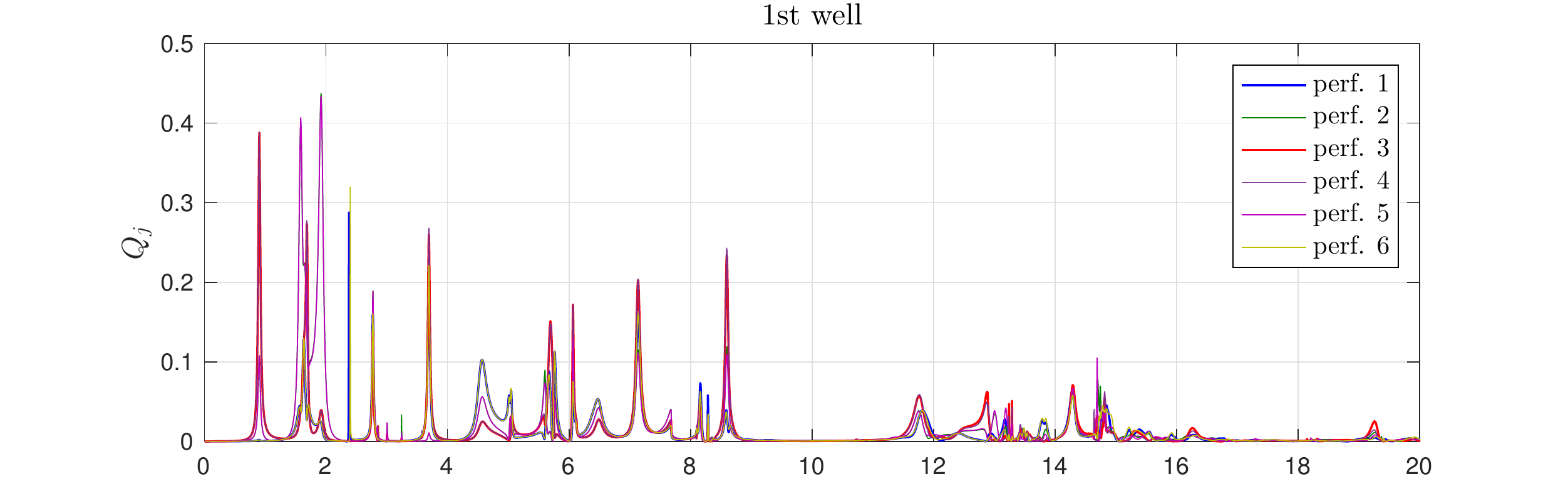}\vspace{-0.1cm}\\
\includegraphics[height=4.5cm]{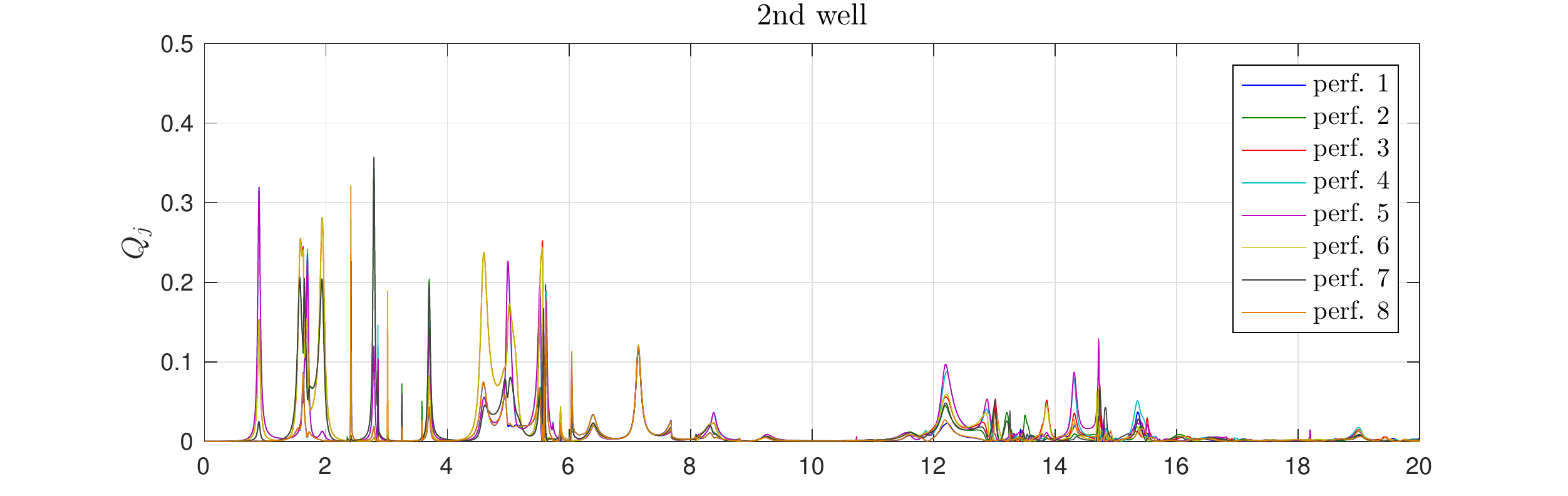}\vspace{-0.1cm}\\
\includegraphics[height=4.5cm]{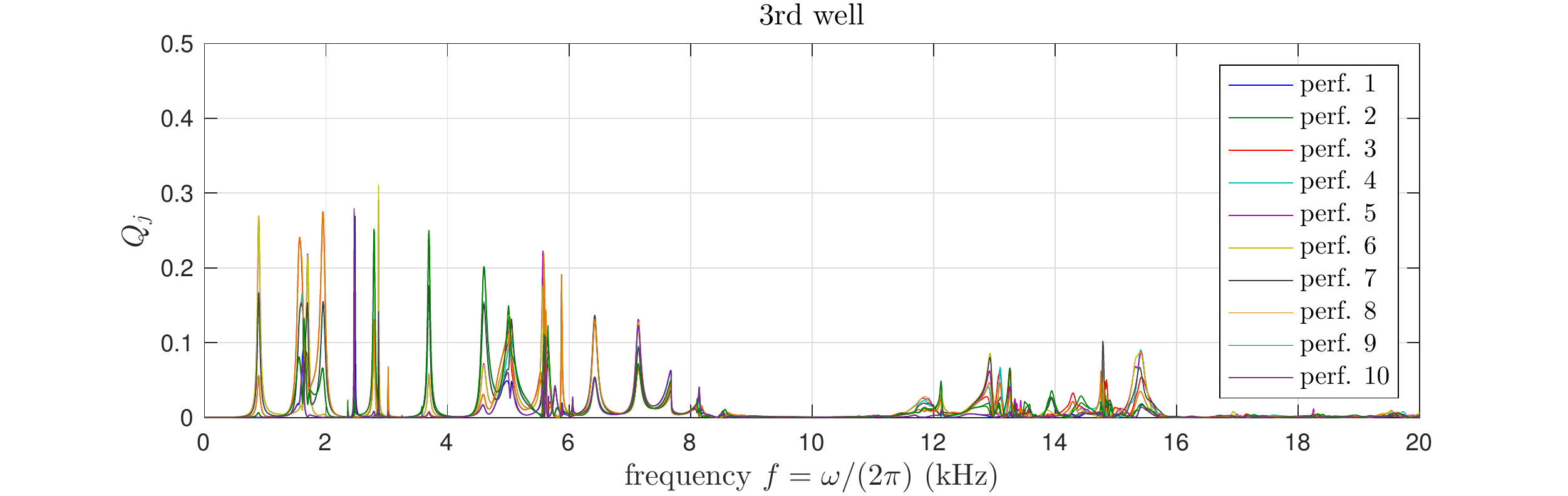}
\end{center}\vspace{-0.5cm}
\caption{Individual energy transmission factors $Q_j$~\eqref{eq:Q_j_factor} as functions of the frequency for the three (1st, 2nd and 3rd) ``symmetric" well configurations displayed in  Figure~\ref{FIG:CH06:WELL_MESH}.}\label{fig:Q_j_factor}
\end{figure}

The large correlation between the location of the peaks for the various well configurations observed in Figure~\ref{fig:Q_factor}, on the other hand, can be explained by the resonant frequencies of an individual perforation. In fact, large values of $Q_j$ are expected to occur at  the resonant frequencies of the $j$-th perforation. Since the same perforation radius, the same perforation length, and the same location of the transducer are utilized in the three configurations considered, all the perforations are expected to resonate collectively at approximately the same frequency. Therefore, the factors $Q_j$, $j=1,\dots,N_p$ attain simultaneously local maxima at these ``resonance" frequencies. To look into that in more detail, we present Figure~\ref{fig:Q_j_factor}---which displays the individual factors $Q_j$, $j=1,\ldots,N_p$---where it can be clearly observed that  the factors $Q_j$ attain collectively local maxima at certain frequencies that indeed correspond to the largest peak values of $Q$ observed in Figure~\ref{fig:Q_factor}. 


\begin{figure}[h!]
\begin{center}
\includegraphics[height=4.5cm]{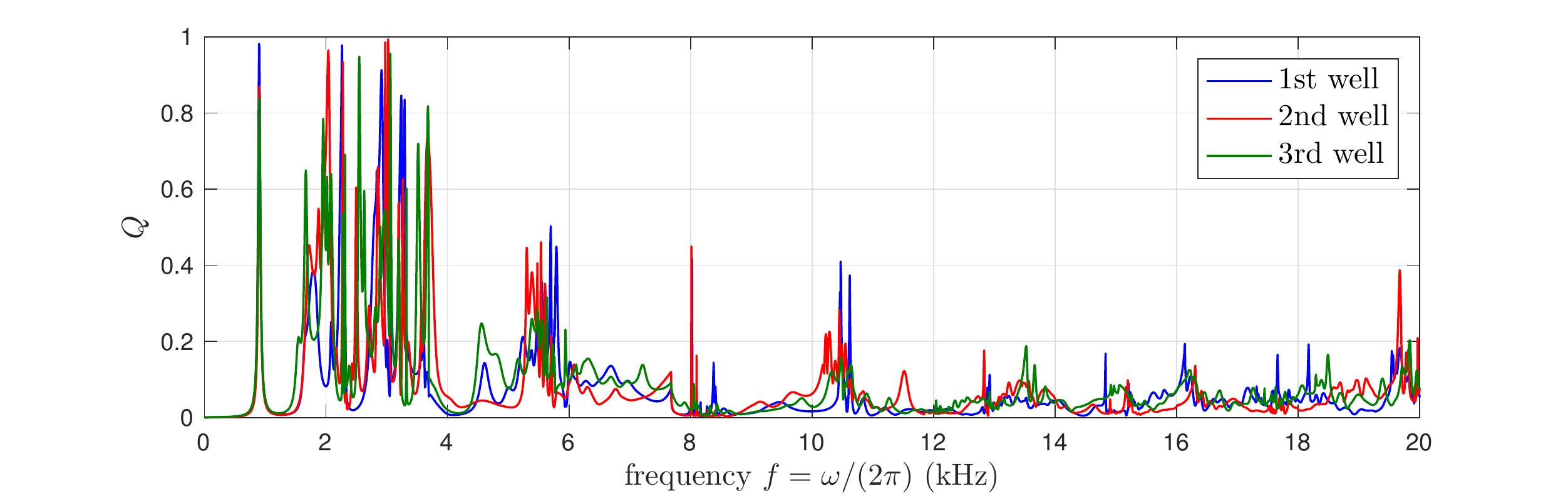}
\end{center}\vspace{-0.5cm}
\caption{Consolidated energy transmission factor $Q$, defined in~\eqref{eq:Q_factor}, as a function of the frequency for  three randomly perturbed well configurations.}\label{fig:Q_factor_rand}
\end{figure}

\begin{figure}[h!]
\begin{center}
\includegraphics[height=4.5cm]{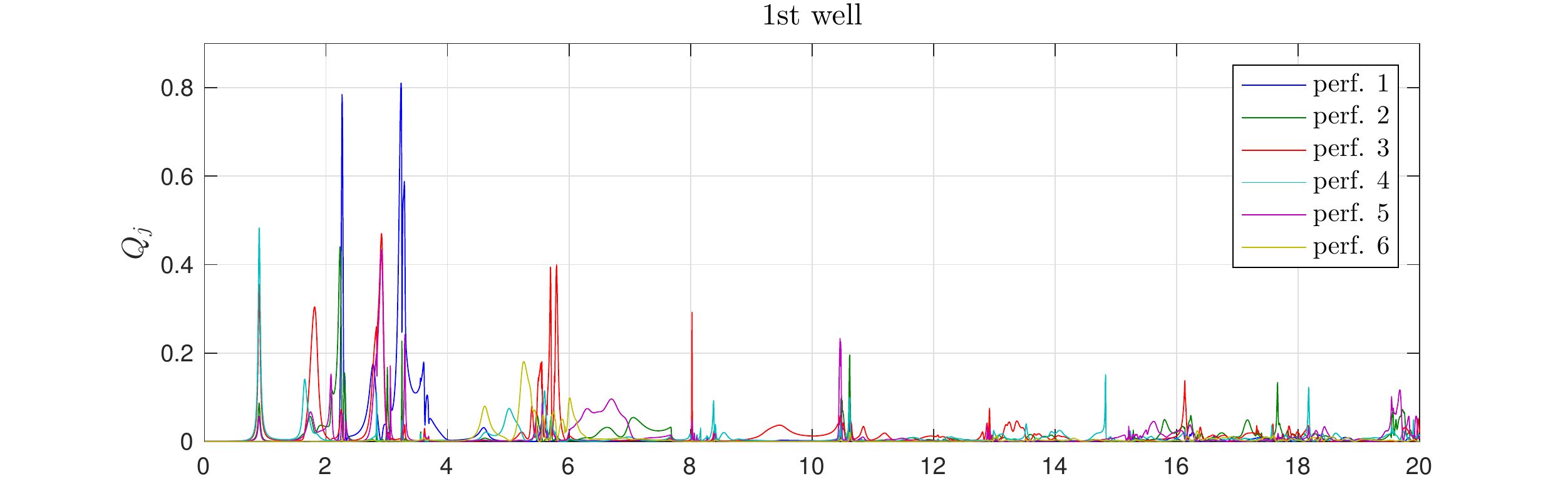}\vspace{-0.1cm}\\
\includegraphics[height=4.5cm]{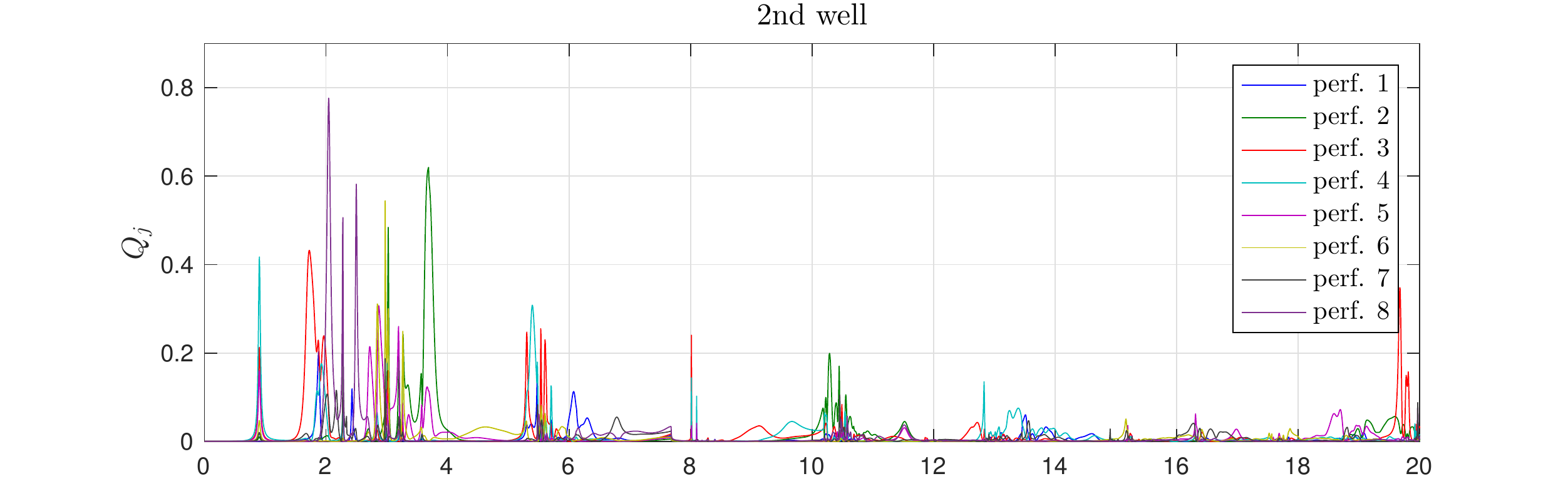}\vspace{-0.1cm}\\
\includegraphics[height=4.5cm]{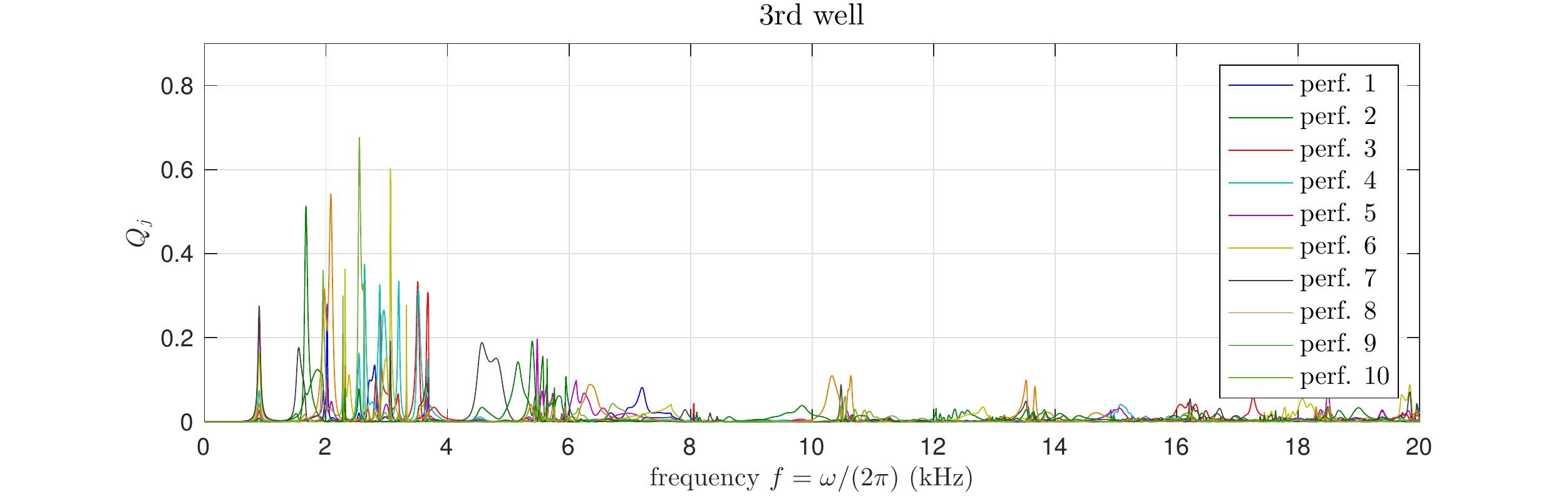}
\end{center}\vspace{-0.5cm}
\caption{Individual energy transmission factors $Q_j$~\eqref{eq:Q_j_factor} as functions of the frequency for  three randomly perturbed well configurations.}\label{fig:Q_j_factor_rand}
\end{figure}

\begin{figure}[h!]
\begin{center}
\includegraphics[width=10.0cm]{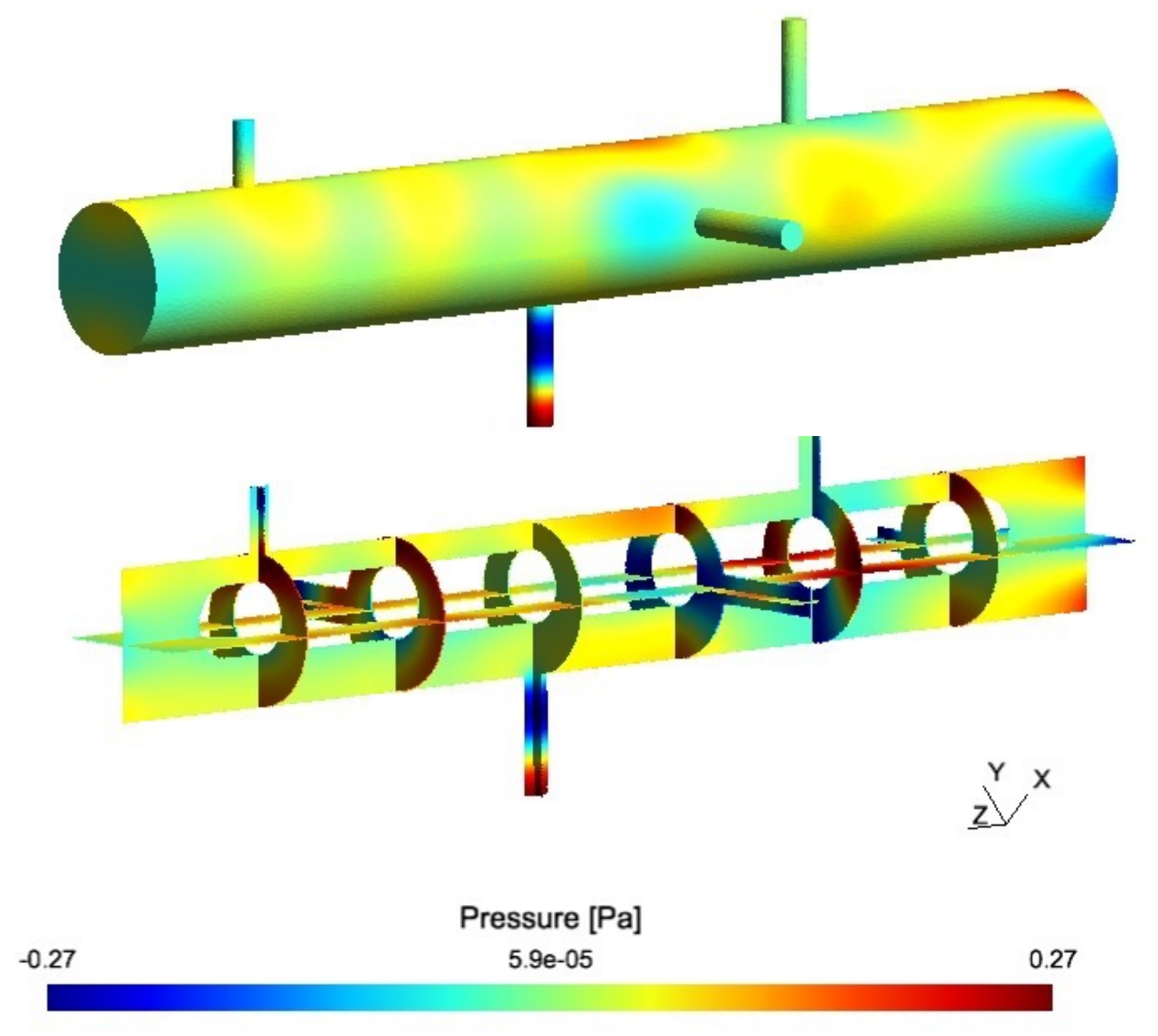}
\end{center}\vspace{-0.5cm}
\caption{Real part of the pressure field inside the perturbed 1st well configuration  at $f=5.7947$ kHz, which corresponds to one of the peak values of the $Q$ factor displayed in Figure~\ref{fig:Q_factor_rand} in blue. Top: pressure field on the boundaries of the computational domain $\Omega$. Bottom: pressure field at various cross sections of $\Omega$. Note the large pressure amplitude values inside the 3rd perforation.}\label{fig:res_2}
\end{figure}

Although the simulation results presented above seemingly indicate the existence of optimal frequencies for which nearly 100\% ($Q\approx 1$) of energy transmission is achieved, in practice, uncertain  variations in the shape of the perforations might result in an overall reduction of  the peak values of~$Q$.  To briefly study the effect of small shape variations on the location of the local maxima of $Q$, we consider perturbations of the three aforementioned configurations, which are generated by introducing random changes in the perforation radius, the perforation length, and the location of the transducer. Figure~\ref{fig:Q_factor_rand} displays the $Q$ factors obtained for the new well configurations, where it can be observed a much weaker correlation between the location of the peak values, as compared to the results presented in Figure~\ref{fig:Q_factor}. This weaker correlation is further explained by the results displayed in Figure~\ref{fig:Q_j_factor_rand}, which show that, as expected, the factors $Q_j$, $j=1,\dots, N_p$ do not attain their local maxima at the same frequencies. Despite this fact, remarkably large peak values of $Q$ ($Q\approx 1$) are still observed. Nearly perfect transmission is achieved in this case by excitation of ``resonant" frequencies associated with just a few perforations, for which the local energy transmission factors $Q_j$ lies well above 50\% (e.g., the plot at the top of Figure~\ref{fig:Q_j_factor_rand}---corresponding to the first well configuration---around $f=3.24$ kHz, where $Q_1=0.81$). Figure~\ref{fig:res_2} displays  the pressure field at one of the peak values of $Q$, where large pressure amplitude values inside some of the perforations are again observed.  We thus finally conclude that, in principle, it would be possible to achieve nearly perfect transmission for realistic well configurations, provided the model assumptions are satisfied.


\section{Concluding remarks}\label{SC05:CONCLUSIONS}
A mathematical model---based upon the Helmholtz equation and the use of a suitable impedance boundary condition---and a DtN-FE procedure are presented for the numerical simulation of an AWS method. The existence of optimal emission frequencies, associated with acoustic resonance phenomena, is demonstrated by means of numerical simulations for a variety of realistic well configurations. We believe that the proposed methodology and the numerical results presented in this work provide valuable information for design and optimization of the AWS method as its performance can be significantly improved by properly selecting the operating frequencies of the AWS device (transducer).

\appendix

\section{White's wall impedance model}\label{app:impedance_bc}

Let us consider a circular cylinder of radius $r_0>0$ which is assumed to be filled with a liquid and surrounded  everywhere by an unbounded porous material. The pressure $P$ and the average flow velocity $V$ in the radial direction are related by Darcy's law
\begin{equation}
\frac{\p P}{\p r}=-\frac\eta\kappa V, \label{eq:darcy}
\end{equation}
where~$\eta$ denotes the shear viscosity of the fluid, and~$\kappa$ denotes the permeability of the porous material. Note that it is assumed in \eqref{eq:darcy} that both the elastic expansion of the tube and the average direction of the flow through the pore space, are radial. The equation of  conservation of mass
\begin{equation*}
 \frac{\p \varrho}{\p t}+\frac1r\frac{\p}{\p r}(r\varrho V)=\frac{\p \varrho}{\p t}+\varrho\left(\frac{\p V}{\p r}+\frac Vr\right)=0,
\end{equation*}
 together with the compressibility relation
\begin{equation*}
B=\varrho\frac{\p P}{\p\rho}=\phi\varrho\left(\frac{\p P}{\p t}\right)/\left(\frac{\p \varrho}{\p t}\right),
\end{equation*}
where $B$ denotes the bulk modulus of the fluid in the pore space and $\phi$ denotes the porosity, lead to
\begin{equation}
\frac{\p V}{\p r}+\frac Vr=-\frac\phi B\frac{\p P}{\p t}.\label{eq:continuity}
\end{equation}
Combining equations~\eqref{eq:darcy} and~\eqref{eq:continuity} we arrive at
\begin{equation*}
\frac{\p^2 P}{\p r^2}+\frac1r\frac{\p P}{\p r}=m\,\frac{\p P}{\p t},
\end{equation*}
where $m=\phi\eta/(\kappa B)$. Further assuming that the velocity and pressure fields in the porous material are time-harmonic, i.e., $P(r,t)={\real}\{p(r)\e^{-i\omega t}\}$ and $V(r,t)={\real}\{v(r)\e^{-i\omega t}\}$, we obtain that the pressure amplitude~$p$ satisfies the Bessel differential equation
\begin{equation*}
\frac{\de^2 p}{\de r^2}(r)+\frac1r\frac{\de p }{\de r}(r)+i\omega m\,p(r)=0,\quad r>r_0.
\end{equation*}
Looking for bounded outgoing-wave solutions at infinity fulfilling the boundary condition $p(r_0)=p_0$ at the interface between the fluid and the porous material ($r=r_0$), we arrive at
\begin{equation}
p(r)=p_0\frac{H^{(1)}_0(\sqrt{i\omega m}\:r)}{H^{(1)}_0(\sqrt{i\omega m}\:r_0)}, \quad r\geq r_0, \label{eq:approx_pressure}
\end{equation}
where $H^{(1)}_0$ denotes the Hankel function of the first kind and order zero~\cite{ABRAMOVITZ:1972}. From Darcy's law~\eqref{eq:darcy}, on the other hand, we obtain that the velocity amplitude $v$ is given by
\begin{equation}
v(r)=\frac{\kappa p_0\sqrt{i\omega m}}{\eta }\frac{H^{(1)}_1(\sqrt{i\omega m}r)}{H^{(1)}_0(\sqrt{i\omega m}r_0)},\quad r\geq r_0.\label{eq:approx_velocity}
\end{equation}
Combining~\eqref{eq:approx_pressure} and~\eqref{eq:approx_velocity}, it is straightforward to evaluate the wall impedance~$Z$, which is defined as the quotient of the pressure amplitude $p$ to the radial velocity amplitude $v$ at the surface of the cylinder, that is,
\begin{equation}
Z(\omega)=\frac{p(r_0)}{v(r_0)}=\Bigg(\frac{\kappa\sqrt{i\omega m}}{\eta}\frac{H^{(1)}_1(\sqrt{i\omega m}r_0)}{H^{(1)}_0(\sqrt{i\omega m}r_0)}\Bigg)^{-1}.
\end{equation}
This wall impedance, given in terms of the frequency $\omega$, accounts for the effect that the porous medium has on the fluid dynamics inside the cylinder.
\bibliographystyle{abbrv}

\bibliography{References}

\end{document}